\documentclass[twocolumn]{revtex4}
\usepackage{amsmath}
\usepackage{amsfonts}
\usepackage{amssymb}
\usepackage{graphicx}

\begin{document}

\title{Anisotropy of Low Energy Direct Photons in Relativistic Heavy Ion
Collisions}
\author{T.\ Koide and T.\ Kodama}
\address{Instituto de F\'{\i}sica, Universidade Federal do Rio de Janeiro, C.P.
68528, 21941-972, Rio de Janeiro, Brazil }

\begin{abstract}
We investigate the behavior of low energy photons radiated by
the deceleration processes of two colliding nuclei in relativistic heavy ion
collisions using the Wigner function approach for electromagnetic radiation fields. 
The angular distribution reveals the information of the
initial geometric configurations. Such a property is reflected in the
anisotropic parameter $v_{2}$, showing an increasing $v_{2}$ as energy
decreases, which is a behavior qualitatively different from $v_{2}$ from
hadrons produced in the collisions.
\end{abstract}

\maketitle

\address{Instituto de F\'{i}sica, Universidade Federal do Rio de Janeiro, C.P.
68528, 21941-972, Rio de Janeiro, Brazil }

\section{Introduction}

In the physics of relativistic heavy ion collisions, the determination of
the initial collision geometry is one of the fundamental pieces to investigate
the dynamics of the created matter such as quark-gluon plasma. Then the
information of the reaction plane is indispensable to study the anisotropic
collective flows of the matter. Such a geometry is deduced indirectly from, 
for example, the statistical average over an event
ensemble by calculating the cumulant of the correlation functions of hadrons which
are generated through very complex strong interactions \cite{v2-ref}. On the
other hand, photons do not suffer from strong interactions, and the
so-called direct photons are considered to carry the information of the early
stage of the collisions. Numerous works in this line have been done from the early days of 
the relativistic heavy ion program \cite{kap1,bjo,lip1,lip2,rus,dmi,eic,jeo,kap2,koc,chj,gal2,gal3,ble,ala,cha,sat}. 
See Ref.\ \cite{PHSD} for a recent review on this subject and references therein.

Among various mechanisms for producing the direct photons, we can consider
bremsstrahlung radiations. This process is usually modeled as classical 
radiations from the decelerated protons of the incident nuclei \cite{kap1,bjo,lip1,lip2,dmi,eic,jeo,kap2,koc,sat}. 
In Ref.\ \cite{Miklos}, the authors focused on the behavior of the higher energy ($\gtrsim 1$ GeV) of the 
photons which are dominantly produced by the incoherent sum of the bremsstrahlung radiations 
from individual decelerated protons, reproducing the spectrum of the observed one \cite{PHENIX}.
In this case, any meaningful information for the initial geometry of
the nuclear scale is expected to be washed out.

On the other hand, for the lower energies, electromagnetic fields may be
generated \textit{coherently} from each decelerating proton, when their
spatial separation is the order of the corresponding wavelength of the
radiations. If this occurs, we expect the following two effects. One is that
the amount of the radiations increases as $\sim Z_{eff}^{2}$, instead of $%
2Z_{eff} $ in the incoherent case, where $Z_{eff}$ is the effective number
of the charges which contribute to the electromagnetic radiations in the
collisions. The other is that the angular distribution of the radiated
photons will reflect the geometric configuration due to the interference of 
radiations from the two incident nuclei. When we have a sufficient yield of the coherent
photons in the very low transverse momentum $p_{T}$ region, the 
elliptic flow $v_{2}$ for the direct photons will be dominated by such coherent 
photons.

In this work, we study the photon spectrum and its angular distributions of
the low energy photons, which are produced by the coherent radiations from two decelerated
incidents nuclei. 
We first calculate the electromagnetic fields by introducing the simplified
trajectories of the two incident nuclei. These nuclei are treated as point-like objects with an effective charge $Z_{eff}$. 
From this, we obtain the phase
space distribution of the photons with the help of the Wigner function which
expresses the photon spectrum and the angular distribution. We further show that the
corresponding anisotropic parameter $v_{2}$ reveals a very enhanced nature in the 
lower $p_{T}$.

In the following, we use $\hbar=c=\varepsilon_{0}=\mu_{0}=1$ and the fine
structure constant is defined by $\alpha_{EM}=e^{2}/(4\pi)$ in the SI (rationalized)
unit.

\section{Model of collisions and electromagnetic radiations}

\begin{figure}[h]
\begin{center}
\includegraphics[scale=0.3]{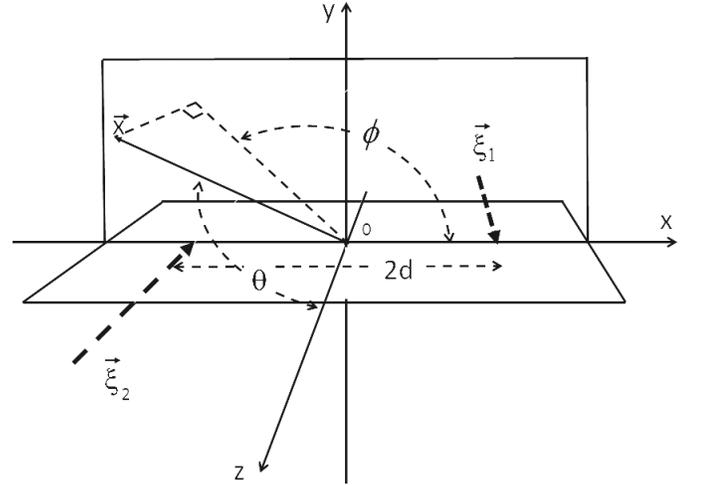}
\end{center}
\caption{The schematic figure for the collision of two incident nuclei.}
\label{collision}
\end{figure}

Let us consider a collision of two identical nuclei with the impact parameter $b$%
. In the strong coherence limit, we can simplify the situation by replacing
these nuclei with point-like particles which have an effective charge $%
Z_{eff}$. One may expect that $Z_{eff}$ is the same as the number of
participant protons $Z_{part}(b)$ from one of nuclei, but more generally,
only a portion of the participant protons can contribute to the coherent
radiations. Then we have the restriction, $Z_{eff}\lesssim Z_{part}\left(
b\right) $. We further consider that the protons in each nucleus will be
completely stopped by the collisions with other protons or neutrons, as is
the initial condition of the Landau hydrodynamic model. 
Our geometrical coordinate is represented in Fig.\ \ref{collision}, where 
the $z-$axis is chosen as the collision direction and the two incident nuclei collide at $t=0$.

In general, the deceleration by the collisions occurs in a finite time period, which is characterized 
by stopping time $\tau _{S}$. For the ultra-relativistic heavy ion
collisions, $\tau _{S}$ is given by the order of the Lorentz contracted
thickness of the projectile, $\tau _{S}\sim R/\gamma $, where $R$ and $%
\gamma $ are the nuclear radius and the Lorentz factor, respectively. For
relativistic limit $\gamma \gg 1$, then the stopping time will be very small. For the sake of
simplicity, we consider the infinitesimal limit of $\tau _{S}.$ In this
case, the deceleration is given by the Dirac delta function in time. See
also the discussion in Appendix \ref{app:finite}.

Then the trajectories of the nuclei $1$ and $2$ are, respectively, expressed
in the Cartesian coordinates as 
\begin{equation}
\vec{\xi}_{1}(t)=\left( 
\begin{array}{c}
d \\ 
0 \\ 
t\ V_{0}\ \theta\left( -t\right)%
\end{array}
\right) ,\ \ {\vec{\xi}}_{2}(t)=\left( 
\begin{array}{c}
-d \\ 
0 \\ 
-t\ V_{0}\theta\left( -t\right) ,%
\end{array}
\right) ,  \label{trajectory}
\end{equation}
where $2d$ represents the transverse distance between the two centers of 
mass of the respective participant protons (See Fig.\ \ref{collision}). This
is usually smaller than the impact parameter, $2d\lesssim b.$ At infinite
distance ($t= -\infty$), the nuclei move with a constant speed $V_{0}$ which
should be less than one.

The solution of the Maxwell equations for these trajectories is given by the
Li\'{e}nard-Wiechert potential \cite{BookEM}. Since we are interested in the 
behaviors of the radiations at the detector position, we drop irrelevant
contributions at infinite distance. Then, the contributions from the charge $%
\vec{\xi}_{1}(t)$ are given by 
\begin{eqnarray*}
\vec{E}_{1}(\vec{x},t) &=&\frac{eV_{0}Z_{eff}}{4\pi }\frac{1}{|\vec{x}-\vec{%
\xi}_{1}(t_{1})|}\frac{1}{1-\vec{\beta}_{1}\cdot \vec{n}_{1}}  \notag \\
&&\times \left\{ \left( 1-\vec{n}_{1}\vec{n}_{1}^{T}\right) \vec{e}%
_{z}\right\} \delta \left( t_{1}\right) , \\
\vec{B}_{1}(\vec{x},t) &=&\vec{n}_{1}\times \vec{E}_{1}(\vec{x},t). 
\end{eqnarray*}%
where $\vec{e}_{z}$ is a unit vector of the $z-$axis and 
\begin{equation*}
\vec{n}_{1}\equiv \frac{\vec{x}-\vec{\xi}_{1}(t_{1})}{\left\vert \vec{x}-%
\vec{\xi}_{1}(t_{1})\right\vert },\ \ \ \ \ \vec{\beta}_{1}\equiv \left. 
\frac{d\vec{\xi}_{1}}{dt}\right\vert _{t=t_{1}}
\end{equation*}%
All these quantities appearing on the right hand sides are evaluated at the
emission time $t_{1}$, defined by the causality equation, 
$|\vec{x}-\vec{\xi}_{1}(t_{1})|=t-t_{1}$.

Therefore, eliminating the emission times, $\vec{E}_{1}(\vec{x},t)$ and $%
\vec{B}_{1}(\vec{x},t)$ are reexpressed as
\begin{subequations} 
\begin{align}
\vec{E}_{1}(\vec{x},t)& =\frac{eV_{0}Z_{eff}}{4\pi }\frac{1}{r_{-}^{3}}%
\left( 
\begin{array}{c}
-\left( x-d\right) z \\ 
-yz \\ 
\left( x-d\right) ^{2}+y^{2}%
\end{array}%
\right) \delta \left( t-r_{-}\right) ,  \\
\vec{B}_{1}(\vec{x},t)& =\frac{eV_{0}Z_{eff}}{4\pi }\frac{1}{r_{-}^{2}}%
\left( 
\begin{array}{c}
y \\ 
-\left( x-d\right) \\ 
0%
\end{array}%
\right) \delta \left( t-r_{-}\right) ,  
\end{align}%
\label{E1B1}
\end{subequations}
where $r_{-}=\sqrt{\left( x-d\right) ^{2}+y^{2}+z^{2}}$.
The corresponding electromagnetic fields from $\vec{\xi}_{2}\left( t\right)$
can be obtained by replacing the two parameters, $\left( d,V_{0}\right) $ by 
$\left(-d, -V_{0}\right)$ in Eq.\ (\ref{E1B1}).

\section{Wigner Function of Electromagnetic Fields}

To extract the spectrum of the photons radiated from the classical
electromagnetic fields, Ref.\ \cite{Miklos} employs an interpretation that
the frequency distribution of the radiation energy 
as the energy distribution of photons 
with the help of Einstein's relation. On the other hand, it is known that
the classical electromagnetic field can be interpreted as the wave function
of the corresponding photons \cite{WignerRefs1,WignerRefs2,WignerRefs3,holland}. 
Here we employ this approach to
calculate the photon angular distribution.

Let us introduce a complex vector function as 
\begin{equation}
\vec{F}=\sqrt{\frac{1}{2}}(\vec{E}+i\vec{B}).
\end{equation}%
Then source-free Maxwell's equations can be reexpressed in a similar form to
the Dirac equation as 
\begin{equation}
i\partial _{t}\vec{F}=-i\left( \vec{T}\cdot \nabla \right) \vec{F},
\end{equation}%
with a constraint,%
\begin{equation}
\nabla \cdot \vec{F}=0,  \label{constraint}
\end{equation}%
where $\vec{T}$ is the spin-1 generator of $O\left( 3\right) $. Equation (%
\ref{constraint}) constrains only the initial condition of $\vec{F}$.
From the definition, one can easily see that
the energy density and the Poynting vector are expressed as $\vec{F}^{\ast
}\cdot \vec{F}$ and $-i\ \vec{F}^{\ast }\times \vec{F}$, respectively. Other
properties of this quantum mechanical interpretation of the vector wave
function, see Ref.\ \cite{WignerRefs3}.

To discuss physical observables measured by a detector at a given
location, it is convenient to introduce the phase-space distribution
function, known as Wigner function. In the present case, we have%
\begin{eqnarray*}
f_{W}(\vec{x},\vec{p},t) &\equiv& \int d^{3}\vec{q}\vec{F}^{\ast}(\vec{x}+%
\vec{q}/2,t) \cdot\vec{F}(\vec{x}-\vec{q}/2,t)e^{i\vec{q}\cdot\vec{p}} 
\notag \\
&=& f_{W}^{\left( E\right) }\left( \vec{x},\vec{p},t\right) +f_{W}^{\left(
B\right) }\left( \vec{x},\vec{p},t\right) .
\end{eqnarray*}
where $f_{W}^{\left( E,B\right) }$ represents the contribution from the
electric (magnetic) field. After some algebra, we find 
\begin{eqnarray}
&& \hspace*{-0.5cm} f_{W}^{\left( E,B\right) }\left( \vec{x},\vec{p}%
,t\right) = G^{\left( E,B\right) }\left( \vec{x}-\vec{d},\vec{p};t\right)
+G^{\left( E,B\right) }\left( \vec{x}+\vec{d},\vec{p};t\right)  \notag \\
&& -2\cos\left( 2\vec{p}\cdot\vec {d}\right) G^{\left( E,B\right) }\left( 
\vec{x},\vec{p};t\right),  \label{WignerE} \\
&& \hspace*{-0.5cm} G^{\left( E,B\right) }\left( \vec{x},\vec{p};t\right)
\equiv\int d^{3}\vec{q}  \notag \\
&& \times \left\{ \vec{F}^{\left( E,B\right) }\left( \vec{x}+\vec {q}%
/2,t\right) \cdot\vec{F}^{\left( E,B\right) }\left( \vec{x}-\vec {q}%
/2,t\right) \right\} e^{i\vec{q}\cdot\vec{p}},  \label{F0}
\end{eqnarray}
and%
\begin{align*}
\vec{F}^{\left( E\right) }\left( \vec{x},t\right) & \equiv\sqrt{\frac {1}{2}}%
\frac{eV_{0}Z_{eff}}{4\pi}\frac{1}{r^{3}}\left( 
\begin{array}{c}
-xz \\ 
-yz \\ 
x^{2}+y^{2}%
\end{array}
\right) \delta\left( t-r\right) , \\
\vec{F}^{\left( B\right) }\left( \vec{x},t\right) & \equiv\sqrt{\frac {1}{2}}%
\frac{eV_{0}Z_{eff}}{4\pi}\frac{1}{r^{2}}\left( 
\begin{array}{c}
y \\ 
-x \\ 
0%
\end{array}
\right) \delta\left( t-r\right) .
\end{align*}
Here, $r=|\vec{x}|$.

As is well-known, the Wigner function does not correspond to the
phase-space distribution, since, in general cases, it can take negative
values. However, as shown below, the large distance behavior guarantees the
non-negativity of the Wigner function. Since $\vec{x}$ and $t$ are macroscopic quantities
associated with the physical measurements by a detector, they are much
larger than the magnitude of $1/p$, where $p$ being the order of MeV $\sim$ GeV. 
Thus, the significant contributions in the $q$ integrals in Eq.(\ref{F0}) 
come from the domain satisfying $q\ll r,t$ due to the exponential factor in the integrands.
Therefore we can safely expand them with respect to $q/r$. 
The integrands contain the product of two delta functions with respect to $t$, 
which is approximately reexpressed as $\delta (t-\sqrt{\left( \vec{x}+\vec{q}%
/2\right) ^{2}})\delta (t-\sqrt{\left( \vec{x}-\vec{q}/2\right) ^{2}})\simeq
\delta \left( t-r\right) \delta \left( q_{//}\right)$, where $q_{//}$ is the
component of $\vec{q}$ parallel to $\vec{x}$.
Then we have 
\begin{align*}
G^{\left( E\right) }\left( \vec{x},\vec{p},t\right) & =C\left( \vec{x}%
\right) \delta \left( t-r\right) \int d^{2}\mathbf{q}_{\bot }e^{i\mathbf{p}%
_{\bot }\cdot \mathbf{q}_{\bot }}+O\left( \frac{1}{r^{3}}\right) \\
& \simeq \left( 2\pi \right) ^{2}C\left( \vec{x}\right) \delta \left(
t-r\right) \delta ^{\left( 2\right) }\left( \vec{p}_{\bot }\right) ,
\end{align*}%
where $\vec{p}_{\bot }$ is the orthogonal component of $\vec{p}$ to $\vec{x}$, and 
\begin{equation*}
C\left( \vec{x}\right) =\frac{\alpha _{EM}}{8\pi }\left( V_{0}Z_{eff}\right)
^{2}\frac{1}{r^{2}}\sin ^{2}\theta .
\end{equation*}%
In the above, $\theta $ is the azimuthal angle of $\vec{x}$ with respect to
the $z$ axis (see Fig.1). For $r\gg d$, we find 
\begin{equation*}
G^{\left( E\right) }\left( \vec{x},\vec{p},t\right) \simeq G^{\left(
B\right) }\left( \vec{x},\vec{p},t\right) ,
\end{equation*}%
and
\begin{equation*}
G^{\left( E,B\right) }\left( \vec{x},\vec{p},t\right) \simeq G^{\left(
E,B\right) }\left( \vec{x}+\vec{d},\vec{p},t\right) \simeq G^{\left(
E,B\right) }\left( \vec{x}-\vec{d},\vec{p},t\right) .
\end{equation*}%
Then the Wigner function is finally expressed as 
\begin{eqnarray}
f_{W}\left( \vec{x},\vec{p},t\right) &\simeq &4\left( 2\pi \right)
^{2}C\left( \vec{x}\right) \left\{ 1-\cos \left( \vec{p}\cdot \vec{d}\right)
\right\} \delta \left( t-r\right)  \notag \\
&&\times 2\frac{1}{p^{2}}\delta ^{\left( 2\right) }\left( \Omega _{\vec{p}%
}-\Omega _{\vec{x}}\right) ,  \label{WignerFinal}
\end{eqnarray}%
where $\Omega _{\vec{p}}$ and $\Omega _{\vec{x}}$ are the solid angles for $\vec{%
p}$ and $\vec{x},$ respectively. The factor $2$ in the last term of the above equation comes
from the two opposite directions contained in $\delta ^{\left( 2\right)
}\left( \vec{p}_{\bot }\right) $. Equation (\ref{WignerFinal}) clearly shows that 
$f_{W}\left( \vec{x},\vec{p},t\right)$ is positive semi-definite.

From this Wigner function, the energy distributions in the $\vec{x}$ and $%
\vec{p}$ spaces are given by 
\begin{eqnarray}
&& \frac{d^{3}\mathcal{E}}{d\vec{x}^{3}}=\frac{1}{\left( 2\pi\right) ^{3}}%
\int d^{3}\vec{p}\ f_{W}\left( \vec{x},\vec{p},t\right), 
\end{eqnarray}
and
\begin{eqnarray}
&& \frac{d^{3}\mathcal{E}}{d\vec{p}^{3}}=\frac{1}{\left( 2\pi\right) ^{3}}%
\int d^{3}\vec{x}\ f_{W}\left( \vec{x},\vec{p},t\right),  \label{dEdp}
\end{eqnarray}
respectively.

\section{Photon Spectrum}

Substituting Eq.\ (\ref{WignerFinal}) into Eq.\ (\ref{dEdp}), 
we obtain the momentum spectrum. Because of the Dirac delta functions in Eq.\ (\ref%
{WignerFinal}), note that this is equivalent to the sum of all energies of the
incoming photons to a detector at $\vec{R}_{D}$, 
\begin{equation}
\frac{d^{3}\mathcal{E}}{d\vec{p}^{3}}=\frac{1}{(2\pi )^{3}}\int dt\int_{\vec{%
\Omega}\in D}d^{2}\vec{\Omega}_{\vec{R}_D} R_{D}^{2}\ f_{W}\left( \vec{R}_{D},\vec{p}%
,t\right) ,  \label{dEd3p}
\end{equation}%
where the integral for the solid angle $\vec{\Omega}_{\vec{R}_D}$ is done within the domain 
$D$ corresponding to the aperture of the detector. In the above, we integrate
all photon energies coming into the detector.
Reexpressing this with the photon number $N$, we have 
\begin{equation}
\frac{d^{3}N }{d\vec{p}_{T}^{2}dy}=\frac{1}{2\pi ^{2}}\frac{%
\alpha _{EM}\left( V_{0}Z_{eff}\right) ^{2}}{p^{2}\cosh ^{2}y}\left\{ 1-\cos
\left( 2\vec{p}\cdot \vec{d}\right) \right\} ,  \label{Angular}
\end{equation}%
where $y$ represents the rapidity. See Appendix \ref{app:rapidity}.
In the following calculations, we choose $V_0 = 1$.

\begin{figure}[tbp]
\centering
\includegraphics[scale=0.3]{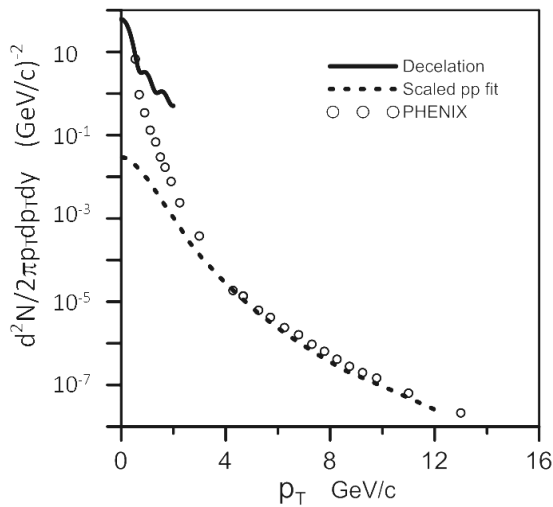}
\caption{The photon spectrum. The solid line represents the results from our
model calculation. The circles and dashed line indicate the PHENIX data \cite{PHENIX} and
the scaled proton-proton collision fit, respectively.}
\label{fig:spectrum}
\end{figure}

For example, let us take $Z_{eff}\sim 80$ and $d\sim 1$ fm as a near central
Au+Au collisions. In this case, the order of the magnitude of the photon spectrum
is%
\begin{equation}
\left. \frac{d^{3}N\left( p\right) }{2\pi p_{T}dp_{T}dy}\right\vert
_{y=0}\simeq 0.37\times 10^{-3}\frac{Z_{eff}^{2}}{p_{T}^{2}}\left(
1-J_{0}\left( 2p_{T}d\right) \right) ,  \label{momentum distribution}
\end{equation}%
where $J_{n}$ is the Bessel function of order $n$. In Fig.\ \ref{fig:spectrum}, 
we show the behavior of the above rough estimate (solid line) together
with the PHENIX data, just for the sake of comparison. Although our
calculation seems to be consistent with the experimental data, 
our idealization of full stopping is not well satisfied in RHIC
energies. Rather, our model will be more suitable for the experiments of the 
lower energies such as NICA or FAIR program \cite{NICA-FAIR1,NICA-FAIR2} where 
the large stopping power is expected. 
Note that if we 
calculate the same spectrum assuming the incoherent radiations as is done in
Ref.\ \cite{Miklos}, the magnitude of the spectrum decreases by one or two
order.

\begin{figure}[tbp]
\centering
\includegraphics[scale=0.25]{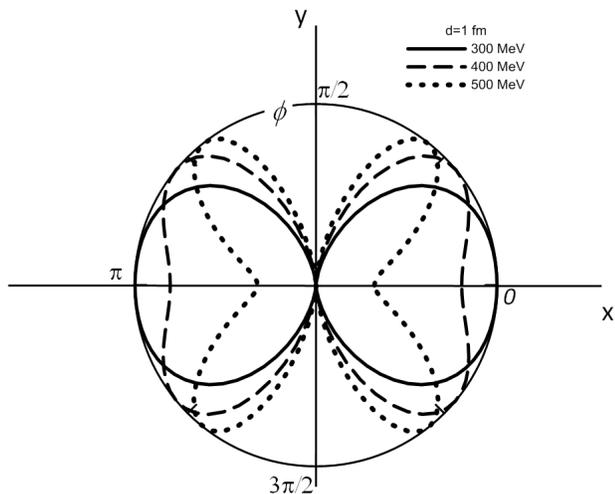}
\caption{The angular distributions at $y=0$. The solid, dashed and dotted
lines represent the results of $p_T = 300$, $400$ and $500$ MeV,
respectively. The axes $x$ and $y$ correspond to those in Fig.\ \protect\ref%
{collision}.}
\label{fig:angular}
\end{figure}

The angular distribution of the photons reveals an interesting behavior as shown
in Fig.\ \ref{fig:angular}. Here, we plotted only the factor 
$1-\cos \left( 2\vec{p}\cdot \vec{d}\right)$ in the radial coordinate with respect
to the azimuthal angle $\phi $ at the vanishing rapidity $y=0$ where $p=p_{T}$, and we find that
there are common dips at $\phi =\pm \pi /2$. These dips correspond to the
direction of the normal vector to the reaction plane. If such a feature is
measurable experimentally, we could determine the event plane unmistakably
and even determine the parameter $d$ quantitatively.

However, unfortunately, the total yield of such low energy photons is very
small $\left( <20\right)$ even in a most favorable condition of our model.
If we consider further experimental difficulties in the detection of the low
energy photons, the determination of the event geometry in the EbyE base seems
to be unrealistic.

\begin{figure}[tbp]
\centering
\includegraphics[
scale=0.3
]{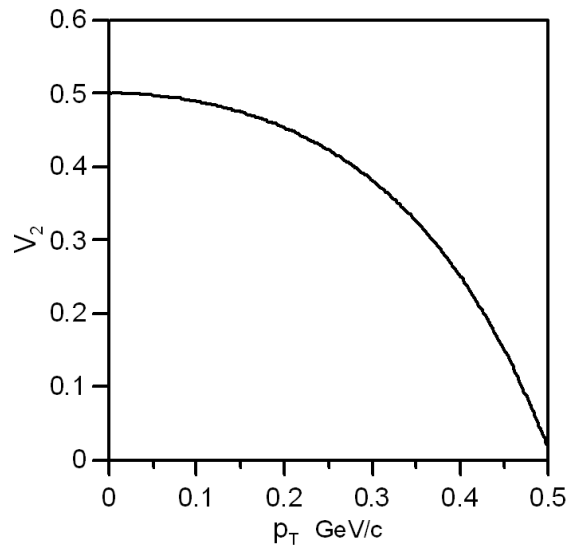}
\caption{Low energy behavior of $v_{2}$ of the direct photons obtained from
the coherent radiations for $d=1$ fm.}
\label{fig:v2}
\end{figure}

On the other hand, the above peculiar behavior will be reflected in another
tractable observables, the anisotropic parameter such as $v_{2}$. In our model, $%
v_{2}$ is calculated as 
\begin{equation}
v_{2}\left( p_{T}\right) =\frac{J_{2}\left( 2p_{T}d\right) }{1-J_{0}\left(
2p_{T}d\right) }.  \label{v2}
\end{equation}%
A similar expression was calculated by Ref.\ \cite{Tamas} in a different context.
In Fig.\ \ref{fig:v2}, we plotted the above $v_{2}$ for $d=1$ fm as before.
In contrast to the well-known behavior of $v_{2}$,
the coherent electromagnetic radiations show an increasing $v_{2}$ for the
lower $p_{T} $ achieving its maximum value $1/2$ for $p_{T}\rightarrow 0$,
independently of the value of $d$. Therefore, if such an increase of $v_{2}$
in the low energy photons $\left( p_{T}< 0.5\ \text{GeV}\right) $ is found
experimentally, it can be considered as the genuine signal from the coherent
electromagnetic radiations by the deceleration, 
although it will be affected by the incoherent radiations. See the discussion in Sec.\ \ref{sec:conc}.
Such a behavior is not expected from the usual hydrodynamic, kinetic or microscopic pictures of the
collective flow mechanism \cite{PHSD,gab}. 

\section{Concluding Remarks and Discussions} \label{sec:conc}

In this short exercise, we investigated the behavior of the low energy
photons radiated by the deceleration processes of the two incident nuclei in
relativistic heavy ion collisions. We assumed that the coherent radiations are 
dominant, so that the two colliding nuclei are replaced by point charges,
and the deceleration mechanism is simply characterized by the Dirac delta
function. We thus consider the full stopping scenario like the Landau type initial condition,
which may have better chance in lower energy heavy ion collisions such as
coming NICA and FAIR experiments \cite{NICA-FAIR1,NICA-FAIR2}.

We found that the angular distribution of the low energy photons reveals
well the initial geometric configurations at the deceleration processes.
Such a property is reflected in the anisotropic parameter $v_{2}$, showing a
very enhanced nature in the lower $p_{T}$. If the angular distribution is measurable 
in the EbyE basis, the initial geometry could be determined. However, the total
photon multiplicity in our model is the order of $10\sim 20$ in an 
optimistic situation, so that the EbyE basis analysis seems to be improbable.
On the other hand, since these signals have characteristic patterns for a
given initial geometry, they may be still useful to improve the
determination of the initial condition by using, for example, the correlations
with other particles. Another interesting possibility for the multiple soft photon 
emission mechanism was suggested in Ref.\cite{Hov}, but the nature of the angular distribution of 
the produced photons would be different from ours.

In this work, we considered a very idealized model of the deceleration where the
coherent electromagnetic radiations occur from the overall nuclear charges,
and did not discuss a mechanism to maintain such a coherence by relativistic heavy ion 
collisions. To clarify these points and examine the above possibilities, it
is important to apply the present approach to more realistic initial
conditions and possible collective deceleration mechanisms, for example, shock wave formation \cite{sat}.  
The Wigner function approach described here will be useful for this purpose. We leave this as a future
task.

\begin{figure}[tbp]
\centering
\includegraphics[scale=0.3]{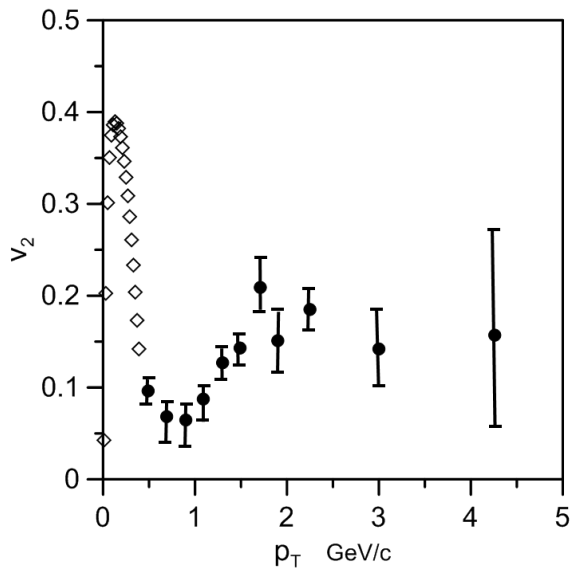}
\caption{$v_{2}$ from direct photons. Squares denote the results with the effects of incoherent
mechanism given by Eq.(\ref{v2-incoherent}) with $\Delta p=0.2$ GeV, Filled circles indicate the experimental
data from PHENIX\cite{PHENIX}.}  
\label{fig:v2-incoherent}
\end{figure}

As shown in Fig.\ \ref{fig:v2}, our coherent radiations of the low energy
photons exhibit an increasing $v_{2}$ as energy decreases, achieving its
maximum value $1/2$ at $p_{T}=0$. This is the case considering only the coherent
radiations. As discussed, the coherence for higher momenta will
be quickly destroyed and the incoherent photons should be dominant. 
We may roughly evaluate such an effect assuming that the coherent
contribution vanishes exponentially with a characteristic scale $\Delta p$
as a function of $p_{T},$ while the incoherent contribution becomes dominant
for $p_{T}\gg \Delta p$. In such a case, the anisotropic parameter $v_{2}$
in Eq.\ (\ref{v2}) is replaced by%
\begin{equation}
v_{2}\left( p_{T}\right) =\frac{J_{2}\left( 2p_{T}d\right)}
{1+2e^{2p_{T}/\Delta p}/Z_{eff}-J_{0}\left( 2p_{T}d\right) }.
\label{v2-incoherent}
\end{equation}%

In Fig.\ \ref{fig:v2-incoherent}, we show the results of Eq.\ (\ref{v2-incoherent}) for $%
\Delta p=0.2$ GeV with squares. For the sake of comparison, the
PHENIX data are plotted together by filled circles \cite{PHENIX}. Note that
in the presence of the incoherent contribution, $v_{2}$ vanishes at $p_{T}=0$
and the maximum is shifted to a finite value of $p_{T}$. Current experimental
measurements of $v_{2}$ of the direct photons are only from $0.5$ GeV and above 
\cite{PHENIX}, and thus it is still difficult to see whether the coherent
radiation mechanism is present or not. However, it is
interesting to note that the experimental data seems to show the beginning
of such an increase for $p_{T} \le 0.5$ GeV as is shown in Fig.\ \ref%
{fig:v2-incoherent}, which is qualitatively in agreement with the behavior
of $v_{2}$ calculated with the coherent radiation mechanism. Of course, our 
deceleration scenario is not applicable to the RHIC experiment, so that any direct 
comparison will not be appropriate. 
On the other hand, if this behavior of $v_{2}$ is attributed to the 
coherent radiations of the photons, we expect that such a signature should be
enhanced in NICA and FAIR. In this aspect, the measurements of the lower energy direct 
photons are essential to clarify the presence of the coherent mechanism in relativistic
heavy ion collisions. 

The authors acknowledge E.\ L.\ Bratkovskaya, G.\ S.\ Denicol, M. Greif and
C.\ Greiner for useful discussions and comments. We also thank E. Kokoulina for calling  our
attention to Ref.\cite{Hov}. This work is financially supported by CNPq and CAPES.

\appendix

\section{Sensitivity of rapidity distribution on deceleration}

\label{app:finite}

The trajectories (\ref{trajectory}) can be
considered as if we take the vanishing $\tau _{S}$ limit of the
parameterization of a continuous deceleration, 
\begin{eqnarray}
\vec{\xi}_{1}(t) &=&\left( 
\begin{array}{c}
d \\ 
0 \\ 
t~V_{0}\ \tanh \left( \frac{t}{\tau _{S}}\right) \ \theta \left( -t\right)%
\end{array}%
\right) , \\
{\vec{\xi}}_{2}(t) &=&\left( 
\begin{array}{c}
-d \\ 
0 \\ 
-t~V_{0}\ \tanh \left( \frac{t}{\tau _{S}}\right) \theta \left( -t\right) ,%
\end{array}%
\right) ,
\end{eqnarray}%
which is similar to Ref.\ \cite{Miklos}, except for the difference in the
argument of $\tanh$.

Substituting this into the above calculations and taking the vanishing limit
of $\tau _{S}$, we find that the factor $\left\{ 1-\cos \left( 2\vec{p}\cdot 
\vec{d}\right) \right\} /p^{2}$ in Eq.\ (\ref{Angular}) is replaced by 
\begin{eqnarray}
&&\frac{1}{2p^{2}}\left\{ \frac{1}{\left( 1-V_{0}\tanh y\right) ^{2}}+\frac{1%
}{\left( 1+V_{0}\tanh y\right) ^{2}}\right.  \notag \\
&&\left. -2\frac{1}{1-V_{0}^{2}\tanh ^{2}y}\cos \left( 2\vec{p}\cdot \vec{d}%
\right) \right\} .
\end{eqnarray}%
In particular, in ultra-relativistic limit $\left( V_{0}\rightarrow 1\right) 
$, the angular distribution of the photons is given by 
\begin{eqnarray}
\frac{d^{3}N }{d\vec{p}_{T}^{2}dy} &=&\frac{1}{2\pi ^{2}}%
\frac{\alpha _{EM}\left( V_{0}Z_{eff}\right) ^{2}}{p^{2}}\left\{ \cosh
\left( 4y\right) -\cos \left( 2\vec{p}\cdot \vec{d}\right) \right\} .  \notag
\label{Continous} \\
&&
\end{eqnarray}%
One can see that the rapidity distribution shows rather hyperbolic increase
for $|y|\gg 1$, so that the photon yield is strongly enhanced in the forward
and backward directions, while the angular distribution tends to be
isotropic. However, for the central rapidity $y=0$, the above result still
coincides with Eq.\ (\ref{Angular}). Therefore, in the plane at the central rapidity, 
the angular distribution of photons is independent of the
deceleration mechanism as far as the time scale $\tau _{S}$ is small enough.
This suggests a possibility that the behavior at $y=0$ is relatively
insensitive for deceleration mechanisms if the characteristic time scale of
the deceleration is enough small.

For the sake of comparison, let us consider the incoherent limit. Then our
spectrum for the deceleration of the Dirac delta function, Eq.\ (\ref%
{Angular}), is replaced by%
\begin{equation}
\frac{d^{3}N }{d\vec{p}_{T}^{2}dy}=\frac{1}{2\pi^{2}}%
\alpha_{EM}V_{0}^{2}Z_{eff}\frac{1}{p^{2}\cosh^{2}y}.  \label{Incoherent 1}
\end{equation}
On the other hand, in the small $\tau_{S}$ limit of the continuous
deceleration, Eq.\ (\ref{Continous}), we have 
\begin{equation}
\frac{d^{3}N }{d\vec{p}_{T}^{2}dy}=\frac{1}{2\pi^{2}}%
\alpha_{EM}V_{0}^{2}Z_{eff}\frac{1}{p^{2}}\cosh\left( 4y\right).
\label{Incoherent 2}
\end{equation}
These rapidity dependences are, respectively, to be compared with the low
energy limit and the Rindler acceleration cases discussed in Ref.\ \cite%
{Miklos}. However, in our case, Eq.\ (\ref{Incoherent 1}) does not
necessarily correspond to the non-relativistic case, since $V_{0}$ can be
arbitrary close to unity, and Eq.\ (\ref{Incoherent 2}) shows more quick
increase in rapidity compared to the large deceleration limit of the Rindler
case. That is, the difference of the deceleration mechanism cganges 
drastically the rapidity distribution of photons. 
See also the related calculations in Ref. \cite{biro2}.

\section{rapidity}

\label{app:rapidity}

Our variables shown in Fig.\ \ref{collision} can be expressed in term of the
rapidity. For the sake of simplicity, we consider the case where the mass is
negligibly small. Then the rapidity is defined by 
\begin{equation}
y=\frac{1}{2}\ln\frac{p+p_{z}}{p-p_{z}}.  \label{app:1}
\end{equation}
Then the energy and longitudinal momentum is expressed as 
\begin{eqnarray}
p &=& p_{T}\cosh y, \\
p_{z} &=& p_{T}\sinh y .
\end{eqnarray}%

On the other hand, we can express $p_z$ as 
\begin{equation}
\cos\theta=\frac{p_{z}}{p}.
\end{equation}
Substituting this into Eq.\ (\ref{app:1}), we have 
\begin{eqnarray}
\cos \theta &=& \tanh y, \\
\sin \theta &=& \frac{1}{\cosh y}.
\end{eqnarray}

\end{document}